\documentclass[runningheads]{llncs}

\usepackage[T1]{fontenc}
\def\doi#1{\href{https://doi.org/\detokenize{#1}}{\url{https://doi.org/\detokenize{#1}}}}

\usepackage{multirow}
\usepackage{tabularx}
\usepackage{amssymb}
\usepackage{amsmath}
\usepackage{bm}
\usepackage{graphicx}
\usepackage{amsfonts}
\usepackage{booktabs}
\usepackage{array}
\usepackage{algorithm} 
\usepackage[algo2e]{algorithm2e} 
\usepackage{algpseudocode}
\usepackage{siunitx}
\usepackage{cite}
\usepackage{hyperref}
\usepackage{mathtools}
\usepackage[misc]{ifsym}

\usepackage{gensymb}
\newcolumntype{C}[1]{>{\centering\let\newline\\\arraybackslash\hspace{0pt}}m{#1}}
 
\usepackage{listings}
\begin{document}
\title{LesionMix: A Lesion-Level Data Augmentation Method for Medical Image Segmentation}
\titlerunning{LesionMix}

\author{Berke Doga Basaran\inst{1,2}\textsuperscript{(\Letter)} \and 
Weitong	Zhang\inst{1} \and
Mengyun Qiao\inst{2,3} \and \\
Bernhard Kainz\inst{1,4} \and
Paul M. Matthews\inst{3,5}\and
Wenjia Bai\inst{1,2,3}}
%

\authorrunning{B. D. Basaran et al.}

\institute{Department of Computing, Imperial College London, London, UK \and
Data Science Institute, Imperial College London, London, UK 
\email{bdb19@imperial.ac.uk} \and
Department of Brain Sciences, Imperial College London, London, UK \and 
Friedrich–Alexander University Erlangen–Nürnberg, DE \and 
UK Dementia Research Institute, Imperial College London, London, UK 
}
\maketitle
\begin{abstract}
Data augmentation has become a de facto component of deep learning-based medical image segmentation methods. Most data augmentation techniques used in medical imaging focus on spatial and intensity transformations to improve the diversity of training images. They are often designed at the image level, augmenting the full image, and do not pay attention to specific abnormalities within the image. Here, we present LesionMix, a novel and simple lesion-aware data augmentation method. It performs augmentation at the lesion level, increasing the diversity of lesion shape, location, intensity and load distribution, and allowing both lesion populating and inpainting. Experiments on different modalities and different lesion datasets, including four brain MR lesion datasets and one liver CT lesion dataset, demonstrate that LesionMix achieves promising performance in lesion image segmentation, outperforming several recent Mix-based data augmentation methods. The code will be released at \href{https://github.com/dogabasaran/lesionmix}{\texttt{https://github.com/dogabasaran/lesionmix}}.

\end{abstract}

\section{Introduction}
Availability of labelled medical imaging data has been a long-term challenge for developing robust machine learning methods for medical image segmentation. In particular, when dealing with lesions or abnormality detection, datasets often follow a long-tail distribution \cite{Galdran2021, Roy2022}, which means there can be a variety of categories for abnormal cases but with each category only containing very few samples. In medical imaging, most data augmentation methods are developed at the image level, aiming to increase the diversity of the full image~\cite{Chlap2021}. They often lack the capability to model specific abnormalities in the images. Recently, several disease-specific augmentation methods have been proposed for brain tumors, multiple sclerosis, and skin lesions \cite{Mok2019, Barile2021, Abdelhalim2021}. Unfortunately, the majority of these methods are either disease or organ-specific, or are difficult to train and implement due to their complexity.

In this work, we propose LesionMix, a novel and simple lesion-level data augmentation method for medical image segmentation. LesionMix is able to populate lesions with various properties, including shape, location, intensity and lesion load, as well as inpaint existing lesions by using a dual-branch iterative 3D framework. With LesionMix, we are able to train lesion segmentation models in a low-data setting, even if there are very few samples of lesion images. We perform a comprehensive evaluation of LesionMix using different imaging modalities and datasets, including four brain MR lesion datasets and one liver CT lesion dataset. Experiments show that LesionMix achieves promising lesion segmentation performance on various datasets and outperforms several state-of-the-art (SOTA) data augmentation methods.

\subsection{Related Works}
\subsubsection{Non-generative data augmentation.} Traditional data augmentation (TDA) techniques are widely used for training medical image segmentation models \cite{Isensee2021}. TDA include flipping, rotating, scaling, intensity changes, and elastic deformations. These augmentations do not dramatically change the lesion properties, such as the shape and location of lesions with respect to the surrounding tissue. Zhang et al. proposed CarveMix, derived from CutMix~\cite{yun2019cutmix}, which uses a lesion-aware Mix-based technique to carve lesion regions from one image and insert them into another image \cite{Zhang2021}. Zhu et al. developed another Mix-based data augmentation method, SelfMix, which performs augmentation by mixing tumours with non-tumour regions~\cite{Zhu2022}. Lebbos et al. introduced semantic mixing for rare lesions in ultrasound images~\cite{lebbos2022adnexal}. Zhang et al. presented ObjectAug, an object-level augmentation method for semantic image segmentation~\cite{Zhang2021ObjectAugOD}. These methods provide valuable insights for lesion-level data augmentation. However, they directly mix original lesion masks for augmentation without augmenting individual lesion volumes, with no attention to location of the augmentation, or the lesion load of augmented images.

\subsubsection{Generative data augmentation.} Generative methods provide an alternative way for data augmentation by performing abnormality synthesis. Salem et al. synthesises multiple sclerosis lesions using an encoder-decoder U-Net structure \cite{Salem2019}. Bissoto, Jin, and Li et al. utilise generative adversarial networks (GANs) to synthesise skin lesions or brain tumours \cite{Bissoto2018, Jin2021, Li2020a}. Reinhold et al. creates lesions with a predetermined lesion load using a structural casual model \cite{Reinhold2021}. Xia et al. employs an adversarial framework for subject-specific pathological to healthy image synthesis, referred to as pseudo-healthy synthesis \cite{Xia2020}. Similarly, Basaran et al. performs lesion image synthesis and pseudo-healthy synthesis by using cyclic attention-based generators \cite{Basaran2022}. Lin et al. proposes InsMix, a data augmentation method for nuclei segmentation, by employing a Copy-Paste-Smooth principle with a smooth-GAN for achieving contextual smoothness \cite{Lin2022}. While generative augmentation methods have potential for diverse abnormality generation, they are often disease-specific and not easy to extend to different applications and datasets. 

\vspace{-.3cm}

\subsection{Contributions} 
There are three main contributions of this work: (1) We propose a novel non-deep data augmentation method, which augments images at the lesion level and accounts for lesion shape, location, intensity as well as load distribution. (2) The method is easy to implement and can be added to a segmentation pipeline to complement traditional data augmentations. (3) It is generic and can be applied to datasets of various modalities (MRI, CT, etc).

\section{Method}
\subsection{LesionMix}
\begin{figure}[h!] \centering
\includegraphics[width=.98\textwidth]{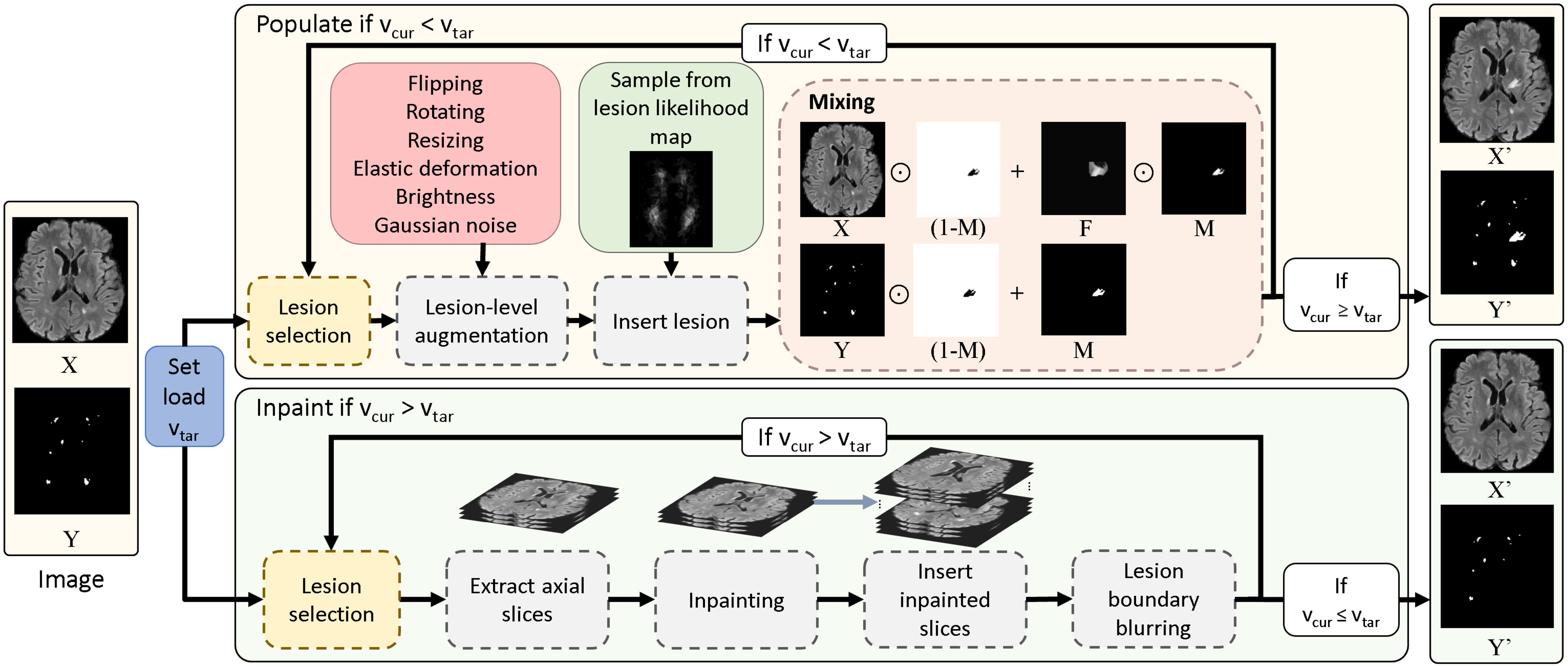}
\vspace{-.2cm}
\caption{Illustration of the augmentation process of LesionMix. It consists of lesion populating (top) and lesion inpainting (bottom) branches to iteratively augment images to a desired lesion load.} 
\label{fig:method}
\end{figure}

The objective is to develop an efficient and easy-to-implement augmentation method that is aware of lesions in medical images, accounting for the spatial and load distribution of the lesions. Figure~\ref{fig:method} illustrates the proposed LesionMix method, which consists of two branches, namely for lesion populating and lesion inpainting. LesionMix takes a lesion image, $\text{X}$, and its corresponding lesion mask, $\text{Y}$, as input, and generates an augmented lesion image, $\text{X'}$, and lesion mask, $\text{Y'}$, as output, which achieves a target lesion load, $\mathrm{v_{tar}}$. If the target lesion load, $\mathrm{v_{tar}}$, is greater than the current load, $\mathrm{v_{cur}}$, the lesion load is increased via the populating branch. Otherwise, the lesion load is decreased via the inpainting branch. To generate diverse lesion samples, lesion-level augmentations are performed during populating. To maintain the fidelity of the samples, lesions are augmented according to learnt spatial and load distributions.
\vspace{-.3cm}

\newpage
\subsection{Lesion populating}
\subsubsection{Lesion-level augmentation.} Given the input image, $\text{X}$, and its lesion mask, $\text{Y}$, a lesion is randomly selected and augmented. We apply 3D spatial augmentations, brightness augmentations (multiplicative), and Gaussian noise augmentations. Lesion-level spatial augmentations include flipping, rotating, resizing, and elastic deformation. Augmentations are applied by extracting the selected 3D lesion volume, applying the augmentation, and inserting the augmented lesion back into the image. By iteratively inserting lesions into the images, augmented lesions can overlap one another, allowing for unique lesion formations. Augmentation parameters are set empirically and provided in Table~\ref{table:augmentations}.

\begin{table*}[h!] \centering
\caption{Data augmentation parameters for LesionMix. $p$ denotes the probability of augmentation. For augmentations where a range is given, the parameter is determined by selecting a value by uniformly sampling from the range.}
\label{table:augmentations}
\begin{tabular*}{\textwidth}{c @{\extracolsep{\fill}} c} \toprule
     {Augmentation} &  {Details}   \\
     \midrule
       Flipping & In X,Y,Z dimensions, $p=0.5$ for each dimension\\ 
       Rotating  & In X,Y,Z dimensions, $p=0.5$ for each dimension, range =[1$\degree$, 89$\degree$] \\
       Resizing & Dimension multiplication, range=[0.5, 1.8]\\ 
       Elastic deformation\footnotemark & Random deformation grid, $\sigma$ range= [3, 7]\\ 
       Brightness & Intensity value multiplication, range = [0.9, 1.1]\\ 
       Gaussian noise & Addition of $\mathcal{N}(0,1)$\\ 
        Inpainting & Fast marching method\\ \bottomrule
\end{tabular*}
\end{table*}

The original image, lesion mask, and the augmented lesion region are mixed using the following equations to generate the augmented image and mask,
\begin{equation}
\label{eq:mix1}
    \text{X'}   =  \text{X} \odot (1-\text{M}) + \text{F} \odot \text{M}
\end{equation}
\begin{equation}
\label{eq:mix2}
    \text{Y'} =  \text{Y} \odot (1 -\text{M}) +  \text{M}, 
\end{equation}
where $\text{F}$ denotes the augmented lesion intensity image, $\text{M}$ denotes the mask of the augmented lesion region, and $\odot$ denotes element-wise multiplication. To generate $\text{X'}$ we use a soft mask $\text{M}$, in which the boundary pixels of the lesion mask are weighted by 0.66 and the inner pixels are weighted by 1. This allows the lesion boundary to blend more naturally with the input image. Lesion populating can be performed iteratively. At each iteration, lesion-level augmentation is applied to a randomly selected lesion and inserted into the image, until the lesion load, $\mathrm{v_{cur}}$, reaches the target lesion load, $\mathrm{v_{tar}}$.

\subsubsection{Lesion likelihood map.} The augmented lesion is inserted into the original image at a location sampled from a spatial heatmap, termed the lesion likelihood map, which describes the probability that a lesion appears at a specific spatial location in the anatomy. The map is learnt by summing the labels of the images to produce a lesion heatmap and normalising it into a probability map. The map is computed once for each organ dataset before model training. For brain datasets, augmented lesions can occur on both the white matter and gray matter. Although white matter lesions may be more common, gray matter lesions have been recorded in clinical literature \cite{Calabrese2013}.

\subsection{Lesion inpainting}
Given the input image and lesion mask, a 3D lesion volume is randomly selected. 2D axial slices of the volume are inpainted using the fast marching method \cite{Telea2004}, which fills in the intensities within the lesion mask with neighbouring intensities from the normal region, formulated by, 
\begin{equation}
\label{eq:fast_marching}
    I(p) = \frac{\sum_{q \in N(p)} w(p,q) [I(q) + \nabla I(q) (p-q)]}{\sum_{q \in N(p)} w(p,q)},
\end{equation}
where $I$ denotes the intensity, $p$ denotes a pixel within the lesion mask, $q \in N(p)$ denotes pixels in the neighbourhood of $q$ that belong to the normal region, $\nabla I(q)$ denotes the image gradient at $q$ and $w(p,q)$ denotes a weighting function determined by the distance and direction from $q$ to $p$ \cite{Telea2004}. After inpainting, we insert the inpainted slices back into the original image. 2D inpainting is implemented on axial slices of the lesions, due to the simplicity in implementation and fast computation. To ensure 3D continuity of the inpainted lesion, Gaussian blurring is performed on the boundary of inpainted lesions along all three dimensions,

\begin{equation}
\label{eq:inpaint1}
    \text{X'} = G(f(\text{X}, \text{M}))\odot \partial \text{M} + f(\text{X}, \text{M}) \odot (1-\partial \text{M})
\end{equation}
\begin{equation}
\label{eq:inpaint2}
    \text{Y'} =  \text{Y} -  \text{M},
\end{equation}
where $f(\text{X}, \text{M})$ denotes the inpainting function using fast marching, $G$ denotes the Gaussian blurring function, and $\partial \text{M}$ denotes the boundary of the lesion mask. Lesion inpainting can be performed iteratively for randomly selected lesions, until the lesion load, $\mathrm{v_{cur}}$, reaches the target lesion load, $\mathrm{v_{tar}}$. 

\subsection{Lesion load distribution}
\begin{figure}[h!] \centering
\includegraphics[width=\textwidth]{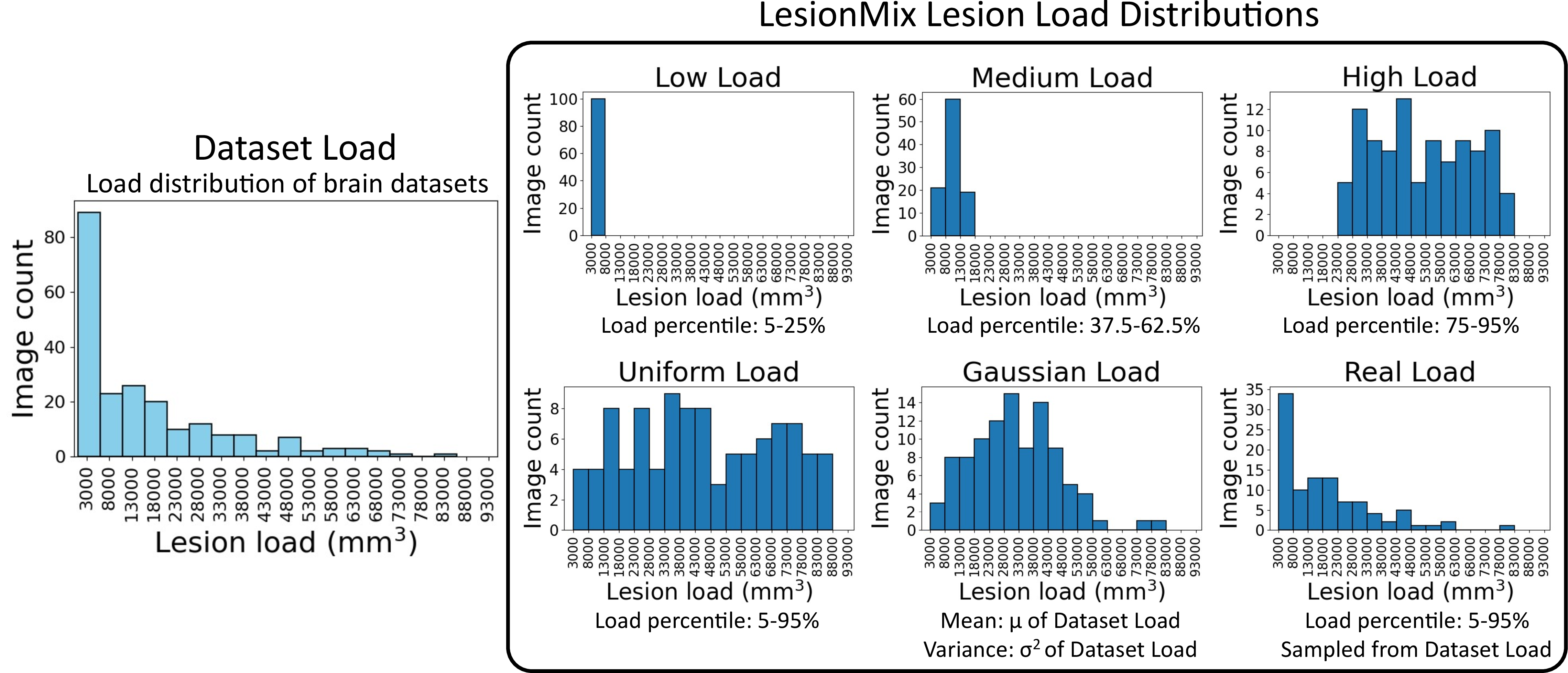}
\caption{Illustration of the brain lesion datasets' lesion load distribution (light blue), named Dataset Load, and the six load distributions (dark blue) for augmentation used by LesionMix. Low load is a uniform distribution sampled between 5 and 25 percentiles of the Dataset Load. Medium load is a uniform distribution sampled between 37.5 and 62.5 percentiles of the Dataset Load. High load is a uniform distribution sampled between 75 and 95 percentiles of the Dataset Load. Uniform load is a uniform distribution sampled between 5 and 95 percentiles of the Dataset Load. Gaussian load samples from a Gaussian distribution with the same mean and variance as the Dataset load. Real load samples directly from the Dataset load distribution. This process is repeated for the LiTS dataset.} 
\label{fig:distributions}
\end{figure}

Unlike other Mix-based methods, LesionMix allows for the generation of datasets with varying lesion load distribution. The load distribution, $P(v)$, is a probability distribution function for lesion volume, $v$. It characterises the degree of severity of the disease. We experiment with six different lesion load distributions: low, medium, high, uniform, Gaussian, and Real. Real denotes the real distribution learnt from the data. The other five are parametric distribution functions, with parameters described in Figure~\ref{fig:distributions}.

\begin{figure}[h!] \centering
\includegraphics[width=\textwidth]{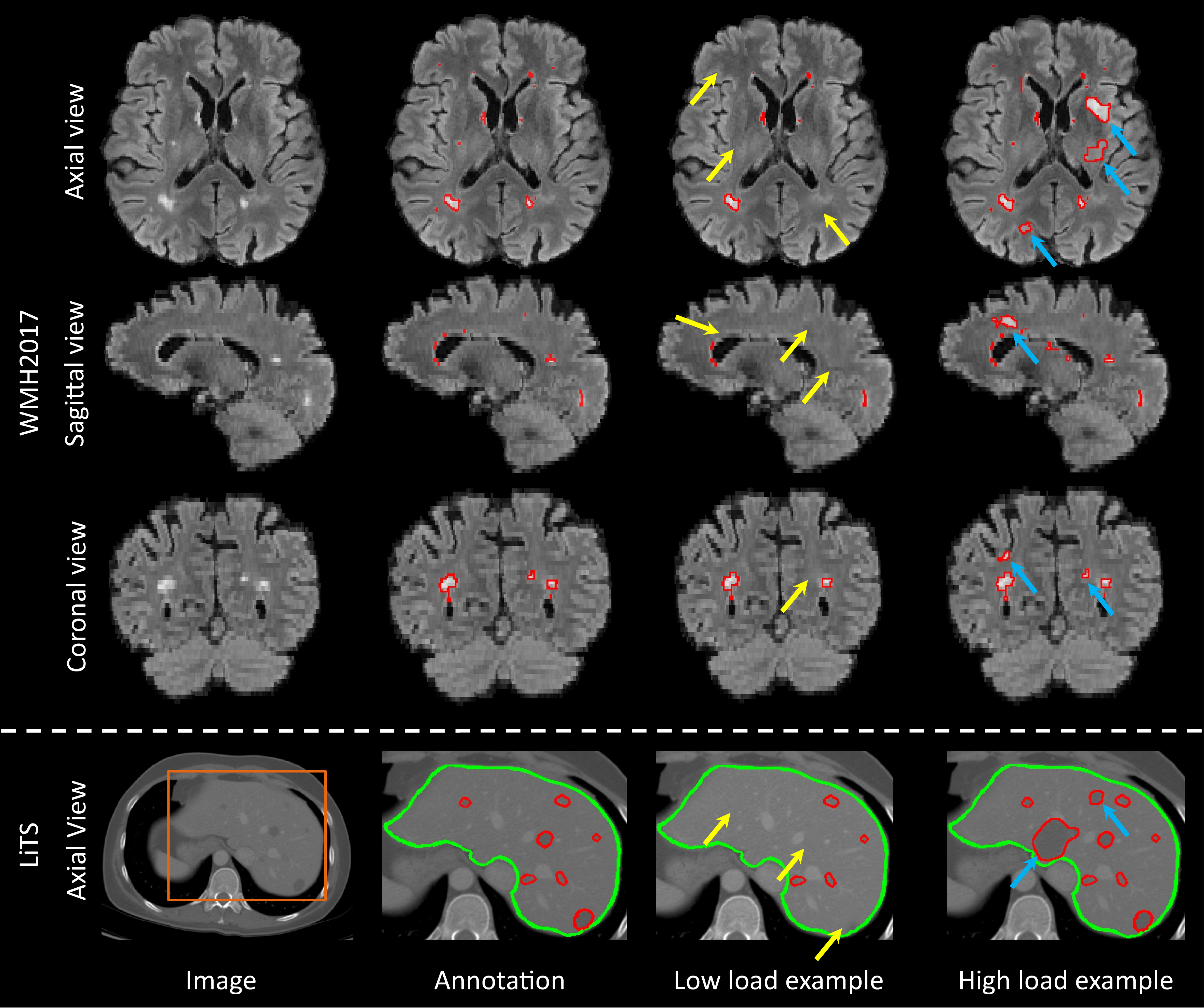}
\caption{Original image, annotation and augmented data by LesionMix with low and high load image examples for both brain (first three rows) and liver (fourth row) datasets. Red denotes brain or liver lesions. Green denotes the liver. Low load example demonstrates performance of inpainted lesions, indicated by yellow arrows. High load example shows populated lesions, indicated by blue arrows.} 
\label{fig:lowhighexample}
\end{figure}

For each image to be augmented, we sample the target lesion load, $\mathrm{v_{tar}}$, from the distribution and apply lesion populating or inpainting iteratively to achieve this target. If $\mathrm{v_{tar}}$ is lower than $\mathrm{v_{cur}}$, lesion inpainting is applied; if $\mathrm{v_{tar}}$ is greater than $\mathrm{v_{cur}}$, lesion populating is applied. We present examples of inpainting and populating as examples of LesionMix in low load and high load distribution setting examples, respectively, in Figure~\ref{fig:lowhighexample}. We present the algorithm of LesionMix in Algorithm~\ref{algo:alg}.

\begin{algorithm}
    \DontPrintSemicolon
    \hspace*{\algorithmicindent} \textbf{Input} Training images and annotations \{$\mathrm{(X_{1}, Y_{1}),...,(X_{N}, Y_{N})}$\}; the desired number of augmented images $T$, the desired load distribution $P(v)$.\\
    \hspace*{\algorithmicindent} \textbf{Output} Augmented training data \{$\mathrm{(X'_{1}, Y'_{1}),...,(X'_{T}, Y'_{T})}$\}
    
    \For{ t=1,2,...,T }{
        Sample target load from the distribution, $\mathrm{v_{tar}} \sim P(v)$\\
        \eIf{$\mathrm{v_{cur}}<\mathrm{v_{tar}}$}{
            \While{$\mathrm{v_{cur}}<\mathrm{v_{tar}}$}{
            1) Randomly select lesion from $\mathrm{(X_{i}, Y_{i})}$\\
            2) Sample lesion location from the lesion likelihood map\\
            3) Apply lesion-level augmentations and generate F and M  \\
            4) Apply mixing in Eq. \ref{eq:mix1} and \ref{eq:mix2}
            }
            }{
            \While{$\mathrm{v_{cur}}>\mathrm{v_{tar}}$}{ 
            1) Randomly select lesion from $\mathrm{(X_{i}, Y_{i})}$ and extract axial slice\\
            2) Apply inpainting in Eq. \ref{eq:inpaint1} and \ref{eq:inpaint2} and reinsert slices
            }
        }
    \textbf{return} $\mathrm{(X'_{i}, Y'_{i})}$\;
    }
    \caption{LesionMix: Lesion-level augmentation}
    \label{algo:alg}
\end{algorithm}

\subsection{Properties of LesionMix}
\begin{figure}[h!] \centering
\includegraphics[width=\textwidth]{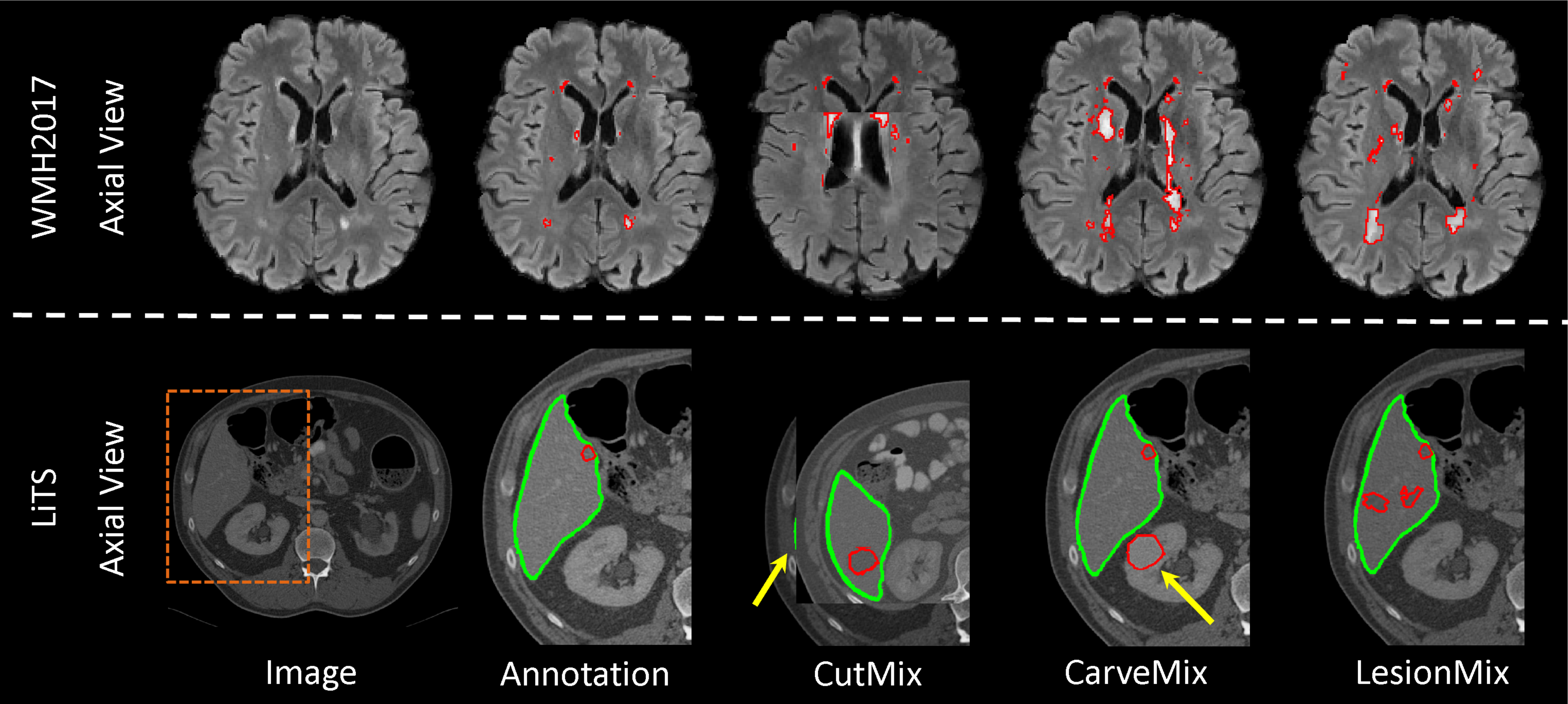}
\caption{Original image, annotation and augmented data by Mix-based methods (lesions: red, liver: green). CutMix produces discontinuities in the image. CarveMix can place lesions outside the organ, indicated by arrows.}  
\label{fig:augs_examples}
\end{figure}

We compare LesionMix with other Mix-based data augmentation methods, including CutMix \cite{yun2019cutmix}, CarveMix \cite{Zhang2021} and SelfMix \cite{Zhu2022}. LesionMix offers greater control in augmentation, compared to CarveMix and SelfMix. LesionMix is spatially-aware, utilising the lesion likelihood map for drawing locations and thus mixes different backgrounds with the lesion. LesionMix performs shape and intensity augmentations at the lesion level, thus increasing sample diversity. Apart from populating lesions, it is also able to inpaint lesions and control the lesion load distribution. We further summarise in Table 2, and present a qualitative comparison against CutMix and CarveMix in Figure~\ref{fig:augs_examples}.

\begin{table}[h!] \centering
\label{table:comparequal}
\caption{Qualitative comparison of the properties of LesionMix with other Mix-based augmentation methods.}
\begin{tabular*}{\textwidth}{l @{\extracolsep{\fill}} lcccc} \toprule
Property   & CutMix [25] & CarveMix [26]& SelfMix [27] & LesionMix \\ \hline
Lesion-aware       &        &   \checkmark       &   \checkmark      &    \checkmark       \\
Spatially-aware   &           &              &                  & \checkmark \\
Lesion-background mixing &   &    & \checkmark       & \checkmark            \\
Lesion-level augmentation       &        &          &      & \checkmark              \\
Lesion inpainting &        &          &         &   \checkmark       \\ 
Lesion load control & & & & \checkmark \\
\bottomrule
\end{tabular*}
\end{table}

\newpage
\section{Experiments}
\subsection{Data}\label{sec:data}
As a generic method for lesion data augmentation, LesionMix is evaluated on brain lesion MR images and liver lesion CT images.  

\subsubsection{Brain lesion data.}
Four brain lesion datasets are used, the MICCAI 2008 multiple sclerosis (MS) lesion dataset (MS2008, $n$=20) \cite{Styner2008}, ISBI 2015 longitudinal MS lesion dataset (MS2015, $n$=21) \cite{Carass2015}, MICCAI 2016 MS lesion dataset (MS2016, $n$=15) \cite{Commowick2018}, and MICCAI 2017 white matter hyperintensity dataset (WMH2017, $n_{train}$=60, $n_{test}$=110) \cite{Kuijf2019}. We use the WMH2017 training set for training a lesion segmentation network with the proposed augmentation method and evaluate its performance on the WMH2017 test set and MS2008, MS2015, MS2016 datasets. For all datasets, FLAIR images are used and resampled to $1\times1\times1$ mm$^3$ voxel spacing, followed by brain extraction using FSL \cite{Jenkinson2012} and rigid registration into the MNI space \cite{Fonov2009}.

\subsubsection{Liver lesion data.}
We use the MICCAI 2017 liver tumor segmentation dataset (LiTS) \cite{Bilic2023}. The training set contains CT scans for 131 subjects, which are split into batch 1 ($n$=28) and batch 2 ($n$=103) by the challenge organisers. We use the LiTS batch 1 dataset for training a liver lesion segmentation network and evaluate its performance on the LiTS batch 2 dataset. The in-plane image resolution ranges from 0.56mm to 1.0mm, and 0.45mm to 6.0mm in slice thickness. The LiTS dataset has high variance of data size and organ shape, therefore we use the normalised label map of the liver as the lesion likelihood map. This ensures the placed lesion is within the liver.

\subsection{Implementation details}
The proposed method is developed on PyTorch. All augmentation methods are evaluated with the same segmentation model, nnU-Net with 3D full resolution configuration, and trained for 1,000 epochs on NVIDIA Tesla T4 GPUs.

\subsection{Results}
\subsubsection{Lesion load distribution.} 
We perform an ablation study to select the optimal lesion load distribution for LesionMix. We simulate a low-data setting by selecting just one training image from WMH2017, perform data augmentation using LesionMix to generate 100 augmented images, and train a segmentation network. Table~\ref{table:lowdata} reports the lesion segmentation performance when six different lesion load distributions are used, and compared against the \textit{None} method, which is trained with a single image without augmentation.

\begin{table*}[h] \centering
\caption{Mean and standard deviations of lesion segmentation Dice scores ($\%$), when different lesion load distributions are used for LesionMix. Best results are in bold.}
\label{table:lowdata}
\scalebox{1}{
\begin{tabular*}{\textwidth}{l @{\extracolsep{\fill}} cccccccc} \toprule
      Test set & None & Low & Medium & High & Uniform & Gaussian & Real \\ 
     \midrule
       MS2008 &  $16.46_{15.43}$ & $22.90_{18.59}$ & $22.23_{17.63}$ & $23.57_{18.72}$ &  $\mathbf{24.07_{19.04}}$ & $22.22_{17.59}$ & $22.66_{17.84}$\\
       MS2015 & $37.58_{16.05}$ & $\mathbf{40.51_{9.05}}$ & $37.09_{12.20}$ & $40.37_{15.64}$ & $39.38_{15.68}$ & $36.64_{14.34}$ & $37.68_{14.21}$\\
       MS2016 &  $24.35_{20.07}$ & $36.75_{21.33}$ & $38.10_{20.37}$ & $49.08_{18.05}$ & $\mathbf{50.72_{18.96}}$ & $41.16_{20.04}$ & $50.32_{16.43}$\\
       WMH2017 & $39.30_{25.16}$ & $49.79_{24.88}$ & $51.59_{23.80}$ & $58.99_{21.26}$ & $\mathbf{59.42_{21.25}}$ & $53.23_{22.85}$ & $55.24_{21.56}$\\ \midrule
       LiTS & $3.34_{4.94}$ & $9.40_{6.25}$ & $12.18_{10.34}$ & $13.33_{5.94}$ & $\mathbf{13.65_{8.02}}$ & $12.88_{7.20}$ & $11.99_{7.82}$\\\bottomrule
\end{tabular*}}
\end{table*}

\subsubsection{Comparison to other data augmentation methods.}
\begin{figure}[h!] \centering
\includegraphics[width=\textwidth]{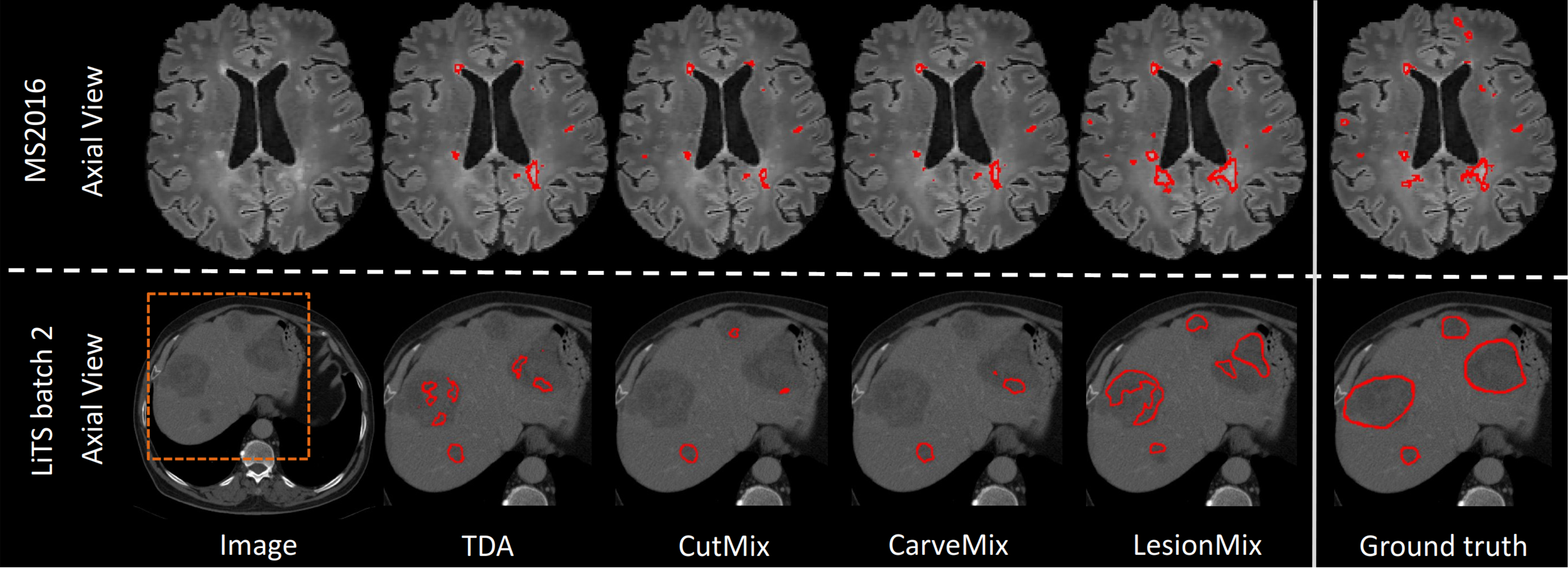}
\caption{Qualitative comparison of segmentation performance when 10\% of dataset size is used. Models with LesionMix detect more lesions and segment them more accurately.} 
\label{fig:segcompare}
\end{figure}

Following the ablation study, we choose the uniform lesion load distribution for the remaining experiments. We compare LesionMix to SOTA data augmentation methods, including traditional data augmentations (TDA), which come default with nnU-Net \cite{Isensee2021}, CutMix \cite{yun2019cutmix} and CarveMix \cite{Zhang2021}. TDA includes rotation, scaling, mirroring, elastic deformation, intensity perturbation and simulation of low resolution. We add CutMix, CarveMix, or the proposed LesionMix onto TDA. We re-implement CutMix \cite{yun2019cutmix} for 3D medical images, and use the public code for CarveMix \cite{Zhang2021}. We are unable to compare against SelfMix \cite{Zhu2022} due to unavailability of public code. For fair comparison, all methods use nnU-net as the segmentation network and augment the WMH2017 training set for brain lesions and the LiTS batch 1 dataset for liver lesions by five times. We conduct experiments when different sizes of the training data is used. Table~\ref{table:segresults} reports the lesion segmentation Dice scores for different data augmentation methods. LesionMix improves lesion segmentation against SOTA methods in the majority of experiments. We notice greater statistical significance in experiments with smaller dataset sizes. We present example segmentations against benchmark methods in Figure~\ref{fig:segcompare}.

\begin{table*}[h] \centering
\caption{Mean and standard deviations of lesion segmentation Dice scores ($\%$), at different sizes of training data. Best results are in bold. Asterisks indicate statistical significance  ($^{*}$:~p$\leq$~0.05, $^{**}$:~p~$\leq$ 0.01, $^{***}$:~p~$\leq$ 0.005) when using a paired Student's \textit{t}-test comparing LesionMix's performance to baseline methods.}
\label{table:segresults}
\scalebox{1}{
\begin{tabular*}{\textwidth}{l @{\extracolsep{\fill}} ccccc} \toprule
           Size & Test set & TDA\cite{Isensee2021} & CutMix\cite{yun2019cutmix} & CarveMix\cite{Zhang2021} & LesionMix \\ 
     \midrule

        & MS2008 &$36.72_{18.07}^{*}$ & $37.67_{19.09}$ & $36.69_{17.82}^{*}$ & $\mathbf{38.30_{17.67}}$ \\
        & MS2015 & $71.59_{11.14}$ & $\mathbf{72.81_{7.53}}$ & $71.33_{9.96}$ & $72.33_{11.41}$\\ 
        100\% & MS2016 & $57.85_{17.50}$ & $63.22_{15.04}$ & $\mathbf{65.85_{14.26}}$ & $65.62_{14.56}$\\ 
        & WMH2017 & $79.13_{10.59}$ & $80.14_{9.69}$ & $79.74_{9.94}$ & $\mathbf{80.95_{9.45}}$\\  
        \cline{2-6}
        & LiTS & $61.93_{24.30}$ & $58.39_{29.40}^{*}$ & $60.20_{29.04}$ & $\mathbf{63.51_{24.97}}$\\ \midrule

        & MS2008 & $35.11_{21.16}$ & $35.83_{19.55}$ & $32.71_{19.95}^{*}$ & $\mathbf{36.04_{18.93}}$\\
        & MS2015 & $70.59_{11.14}$  & $71.77_{7.16}$ & $67.44_{9.78}^{**}$ & $\mathbf{71.82_{7.40}}$\\ 
        50\% & MS2016 & $55.75_{17.82}^{***}$ & $57.90_{14.24}^{**}$ & $62.27_{16.64}$ & $\mathbf{62.64_{16.26}}$\\ 
        & WMH2017 & $73.65_{17.30}$ & $74.26_{12.69}$ & $72.45_{18.25}^{*}$ & $\mathbf{75.65_{17.60}}$\\ 
        \cline{2-6}
        & LiTS & $52.40_{29.21}$ & $49.60_{31.32}^{*}$ & $51.98_{27.99}$ & $\mathbf{52.59_{29.18}}$\\ \midrule

        & MS2008 & $34.35_{21.49}$ & $33.86_{20.21}$ & $30.64_{21.17}^{*}$ & $\mathbf{34.51_{20.04}}$\\
        & MS2015 & $66.09_{8.82}^{**}$ & $67.69_{6.10}^{*}$ & $67.12_{8.13}^{*}$ & $\mathbf{70.73_{8.20}}$\\ 
        25\% & MS2016 & $55.72_{18.78}^{**}$ & $58.24_{15.67}$ & $60.64_{17.67}$ & $\mathbf{60.94_{16.66}}$\\ 
        & WMH2017 & $71.50_{17.88}^{*}$ & $72.02_{14.29}$ & $72.12_{16.95}$ & $\mathbf{73.92_{15.81}}$\\ 
        \cline{2-6} 
        & LiTS & $29.88_{26.52}^{***}$ & $36.35_{30.12}^{*}$ & $36.77_{29.45}$ & $\mathbf{39.25_{30.01}}$\\ \midrule
        
        & MS2008 & $28.77_{17.77}^{**}$ & $31.32_{18.28}$ & $29.57_{21.66}^{*}$ & $\mathbf{32.28_{22.11}}$\\ 
        & MS2015 & $63.97_{9.97}^{**}$ & $64.97_{10.15}^{*}$ & $58.10_{13.35}^{***}$ & $\mathbf{67.13_{10.73}}$\\ 
        10\% & MS2016 & $43.07_{17.03}^{***}$ & $49.58_{13.65}^{***}$ & $50.34_{13.71}^{***}$ & $\mathbf{61.02_{15.56}}$\\ 
        & WMH2017 & $71.16_{16.70}$ & $70.18_{14.79}^{*}$ & $67.80_{19.97}^{**}$ & $\mathbf{72.03_{15.48}}$\\ 
        \cline{2-6} 
        & LiTS & $24.23_{25.76}^{***}$ & $27.99_{29.26}^{*}$ & $20.53_{26.84}^{***}$ & $\mathbf{30.12_{27.47}}$\\ \bottomrule    
\end{tabular*}}
\end{table*}

\newpage
\section{Conclusion} 
We present LesionMix, a simple lesion-level data augmentation method. It is aware of the lesion likelihood distribution and produces augmented data with varying lesion load. LesionMix improves segmentation performance against other Mix-based augmentation methods across datasets of different modalities and organs. It is modality- and organ-agnostic and can serve as a useful tool for medical image segmentation.

\subsubsection{Acknowledgements} This work is supported by the UKRI CDT in AI for Healthcare \href{http://ai4health.io}{http://ai4health.io} (Grant No. EP/S023283/1).

%
%
\newpage
\newpage
\bibliographystyle{splncs04}
\bibliography{refs}

\end{document}